\documentstyle[prl,aps,multicol,graphicx]{revtex}
\begin{document}
\draft
\title{Non-Crystalline Structures of Ultra-Thin Unsupported Nanowires}
\author{
O\u{g}uz G{\"u}lseren$^{1,2,3,4}$, Furio Ercolessi$^{2,3}$,
and Erio Tosatti$^{1,2,3}$}
\address{$^1$ International Centre for Theoretical Physics (ICTP), Trieste, 
Italy}
\address{$^2$ International School for Advanced Studies (SISSA),
Trieste, Italy}
\address{$^3$ Istituto Nazionale di Fisica della Materia, 
Unit\`a Trieste SISSA}
\address{$^4$ School of Physics, University of Bath, Bath, United Kingdom.
}
\date{23 January 1998 (revised 5 March 1998)}
\maketitle
\begin{center}
Preprint SISSA {\bf 9/98/CM/SS}\\
To appear on {\sl Physical Review Letters}
\end{center}

\begin{abstract}
Computer simulations suggest that ultrathin metal wires should develop
exotic, non-crystalline stable atomic structures, once their
diameter decreases below a critical size of the order of a few atomic
spacings. The new structures, whose details depend upon
the material and the wire thickness, may be dominated
by icosahedral packings. Helical,
spiral-structured wires with multi-atom pitches are also predicted.
The phenomenon, analogous to the appearance of icosahedral and
other non-crystalline shapes in small clusters, can be rationalized
in terms of surface energy anisotropy and optimal packing.
\end{abstract}

\pacs{PACS numbers: 61.46.+w, 61.43.Bn, 81.05.Ys}

\begin{multicols}{2}

The regular crystalline packing of atoms,
which usually minimizes energy in a bulk solid,
is known not to last indefinitely as physical size is decreased.
For very small elemental clusters, for example, one expects
a structural switch from bulk-like to some new arrangement
at some critical radius \cite{ino}.
Such is the case, for example, of small 
(diameter $\sim$ 20--40 \AA{}) clusters 
of Ag \cite{hall} and other metals \cite{marks},
and of rare gas clusters \cite{miehle,raoult}, which were 
observed to abandon single crystal structures in favor of
an icosahedral shape below a critical size.
Cluster calculations based on the Lennard-Jones (LJ) potential 
\cite{raoult,LJclusters} confirm the stability of
the icosahedral shape, implying
a radical deviation from the bulk fcc morphology, for $N\lesssim 4000$.
The driving force causing this morphological change at small size
can generally be traced back to two very distinct sources: 1) electronic
magic sizes, stabilized by filling of shells, as in atoms
and nuclei; 2) competition between optimal internal
packing and minimal surface energy, the latter
dominating for sufficiently small size. Here we shall address
exclusively phenomena caused by the second mechanism, which
is important in a large class of cases. For example, very small 
icosahedral clusters, rather common in nature, are favored by their
ability to expose an exceptional 100\% of (111) faces, 
lowest in energy.

We address the same structural question for ultra-thin nanowires.
{\it A priori}, 
one could anticipate a weaker tendency to depart from the bulk
structure than in clusters, because of the smaller relative 
surface/bulk ratio in the wires
($2/R$ against $3/R$, where $R$ is the radius).
However, we present here detailed optimization results, suggesting
that new non-crystalline nanowire structures (which we propose,
for convenience, to name  {\em weird} wires) should in fact be 
rather readily realized in nature. As it turns out, the tendency for
the wire structure to switch from crystalline to weird appears
to be both general, and stronger than anticipated. 
Moreover, there seems to be a variety of
new structures, changing continuously and unpredictably
from size to size, and from a substance to another. 
For our initial analysis we have chosen to study elemental (unsupported)
metal wires, whose fabrication
should be within reach, thanks to the metal's combined
ductility and strength.

We considered so far Al and Pb as test cases.
Both  were assumed to be described by classical, empirical many-body 
interatomic potentials, of the ``glue''type \cite{glue}.
The glue parameters were previously optimized to model 
the properties of metallic Al \cite{ercadams}
and Pb \cite{lim} respectively. To be sure, such potentials are not 
meant to provide a quantitative description for metal wires
of nearly atomic thickness, whose electronic structure may and will differ
profoundly from that of the bulk metal. The structures thus generated
should therefore be considered at this stage of mostly qualitative 
significance, while future {\em ab initio} studies will be called for 
to achieve quantitative accuracy, and to include electronic shell effects,
expected to be very important at small radii.
Nonetheless, the existence of stable weird wire structures, suggested 
by our energy optimizations with classical potentials, 
represents in our view a 
true, general, and parameter-independent phenomenon. For instance, it is 
found that even a hypothetical argon wire (as described by 
a Lennard-Jones potential) turns weird when sufficiently thin 
\cite{lazzeri}.
We also note that the simplicity of the empirical potentials  for metals
has been at this initial stage quite
instrumental, as it allowed a broad and thorough search for optimal 
structures, which would be prohibitive, had we adopted
from the outset the more accurate {\em ab initio} approach.

\begin{figure}[t]

   \centering 
   \includegraphics[width=3in]{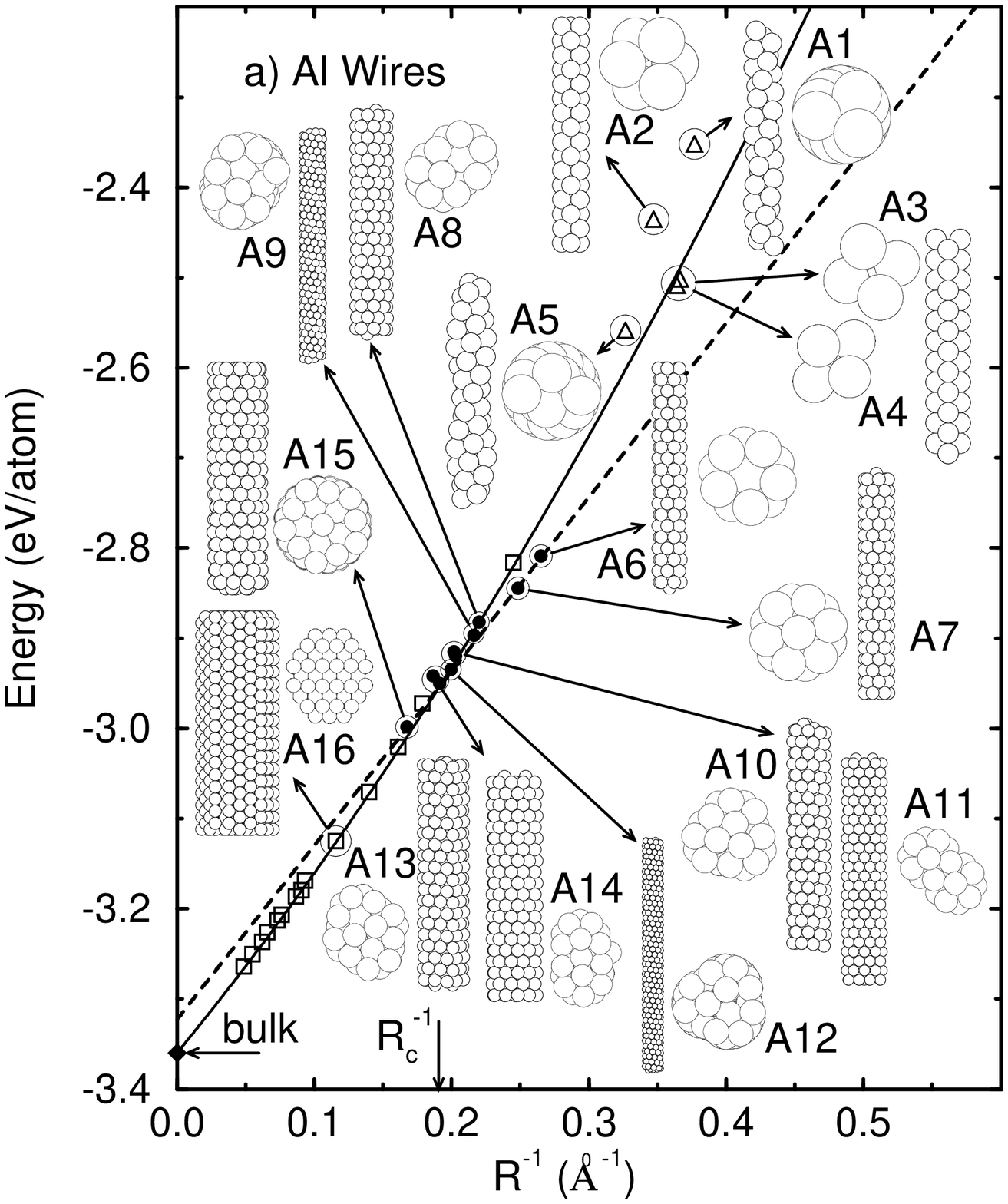}
\hspace{0.5in}
   \includegraphics[width=3in]{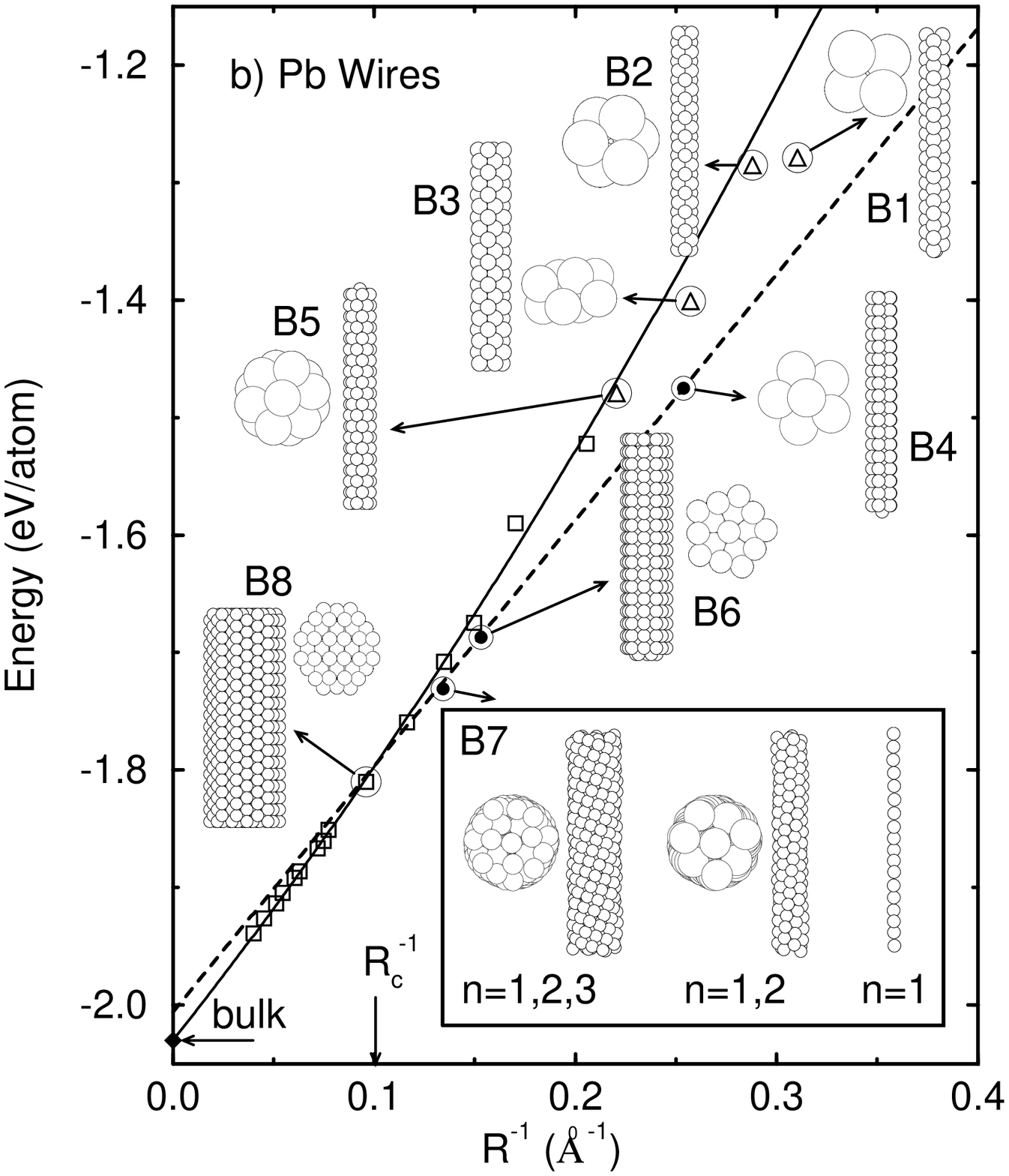}
%
\vspace*{0.5cm}
\caption{
Total energy per atom $E$ vs.\ inverse wire radius $1/R$
for the relaxed structures obtained by optimization 
for Al (a) and Pb (b) wires. A selection of morphologies is shown.
fcc wires are represented by open squares, ``weird'' wires by full circles.
Very thin wires which do not belong to either class
have been marked with open triangles.
Solid lines represent a fit to fcc wires using 
eq.\ (\ref{eq:ecryst}), and dashed lines a fit to weird wires
using (\ref{eq:eweird}).
Weird structures become favored for $R<R_c$.
Inset: structure of the helical Pb wire B7:
complete wire (\protect\mbox{$n=1,2,3$}),
with outer shell removed (\protect\mbox{$n=1,2$}),
and inner strand (\protect\mbox{$n=1$}).
Note how the outer shell exhibits a nearly-square atomic structure,
while that of the second shell is nearly-triangular.
Also note the different helical pitches of these two shells.
}
\label{fig:EvsR}
\end{figure}

For structural optimization, we adopted a 
molecular-dynamics based ``simulated annealing''
methodology, using techniques
previously employed to investigate premelting of thin wires \cite{wiremelt}.
We generally started with fcc (110)-oriented $N$-atom wires
[in a few cases we also used (100) wires, with similar outcomes],
with periodic boundary conditions (PBC) along the wire axis direction $z$.
We first relaxed the atomic positions, to optimize the fcc wire structure.
Starting from this structure, we simulated thermal annealing cycles,
with $T$ rising in steps up to a maximum (typically about
$0.75\,T_m$, where $T_m$ is the bulk melting point)
and then decreasing back to $T=0$.
The wire length was clamped by the PBCs, and its value kept constant 
during annealing, thus preventing surface-free-energy-driven 
contraction into a drop, which thermal diffusion 
would otherwise inevitably drive.
Once back at $T=0$, however, the length was finally allowed to adjust, 
and the wire structure further relaxed, before inspection of 
its energy and structure.
The resulting annealed wire energy was almost invariably better 
than that of the fully relaxed original fcc structure.  Its shape 
generally exhibited large irregularities and a $z$-dependent thickness, 
occasionally with sections showing a new regular structure, which initially 
triggered our curiosity.
Simple annealing was generally unable to generate a fully regular wire.
In a generality of cases where the spontaneous regular sections could be 
identified by visual inspection, new full-length entirely regular wires 
were produced with just that structure. In the new artificial wire, the total 
length and atom number were chosen such as to guarantee 
a proper matching through PBCs. After further relaxation we obtained 
the final wires, collectively depicted in fig.\ \ref{fig:EvsR}.

\end{multicols}

\begin{multicols}{2}

The wire radius $R$ is defined by $S=\pi R^2$, where $S$
is the area projected on the $xy$ plane
by all the atoms (considered as spheres of diameter $d$, where $d$ is
the nearest-neighbor distance in the bulk crystal)
included in a portion of the wire of length 
$2d \sqrt{2/3}$, corresponding to two (111) interplanar spacings.

Crystalline wires clearly prevail for $R$ larger than a critical value $R_c$
of the order of $3d$, depending on the system.
For $R<R_c$, weird structures make their appearance.
Their energies fall systematically {\em below} the extrapolated
fcc line, consistent with simulations, showing an irreversible and exothermic
spontaneous restructuring of the initially fcc wire.
The structures found for Al and Pb
are found to bear certain resemblances, but are certainly not identical,
although we have not attempted a complete search. A (simplified) 
visual illustration of some stable weird wire morphologies
is offered in Fig.\ \ref{fig:EvsR}.
These structures, though noncrystalline, are very regular.
The thicker ones, about four atomic radii in diameter, are three-shell
wires (A15, B6, B7), where a central string is surrounded
by two successive coaxial cylindrical shells, mutually
related by some kind of epitaxy.
Several structures, including three-shell (A15, B6)
and two-shell (A6, B4) cases
and several composites (A8--A14), display pentagonal motives. 

\begin{figure}
\includegraphics[width=2.3in,angle=270]{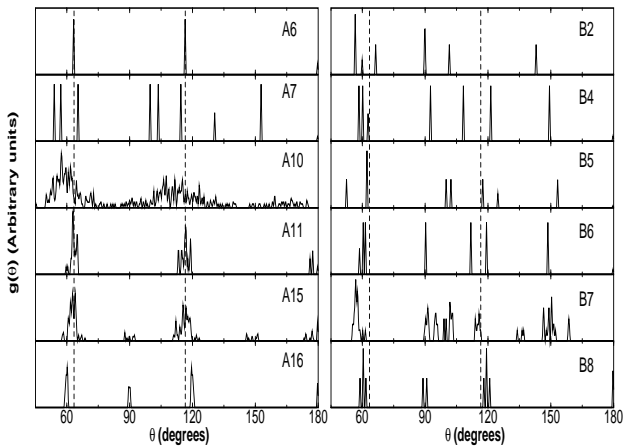}
\vspace*{0.5cm}
\narrowtext
\caption{
Angular correlation functions, restricted to corner atoms located
at or near the wire axis.
Al wires are on the left panel, and Pb wires on the right panel.
Dashed lines correspond to perfect icosahedral angles
($63.4^\circ$, $116.6^\circ$ and $180^\circ$).
A16 and B8 represent a nearly perfect fcc packing,
and A6 is a nice example of icosahedral order.
}
\label{fig:angular}
\end{figure}

For better characterization, we have calculated
angular correlation functions for all our structures.
These are defined by considering all the angles formed by bonds
$ij$ and $ik$, where $i$ runs on all the atoms closest to the wire
axis, and $j$ and $k$ on their neighbors.
The cutoff distance used to define the neighbors is 
$3.3\,\rm\AA$ for Al and $3.5\,\rm\AA$ for Pb.
Results for a selection of cases are shown in 
fig.\ \ref{fig:angular}.
Wires A16 and B8 exhibit peaks at 60, 90, 120 and 180 degrees
typical of regular fcc structure.
In contrast, the pattern of the pentagonal wire A6 has only three 
peaks near 63, 117 and 180 degrees, the signature of icosahedral
packing.  Strong icosahedral features can be also observed on
A11 and A15.   A local icosahedral geometry has recently been found
in a Na neck studied by first principles molecular dynamics \cite{landman}.
Occasionally we find structures like A10 exhibiting
a broad distribution, indicating a marginal case.
In the thinnest wires (probably less meaningful than the others),
we also observe triangular  and tetrahedral packings.
Pb wires also exhibit pentagonal motifs (B4, B6), but angles show
no icosahedral packing: this different behavior is discussed below.

Some wires, such as A1, A5, A9, A12, B7, 
are truly weird, possessing chiral, or helical, arrangements, with 
long periods. In these cases
we also checked for stability against a discrete change of pitch,
although we could not generally rule out the possibility of
a continuously changing (incommensurate) pitch. 

The detailed internal
structure of a beautiful weird wire, the three-shell helical
B7 of Pb, is illustrated in Fig.\ \ref{fig:EvsR}b (inset).
The center ($n=1$) consists of a single, nearly straight atomic chain, 
or strand.
The surrounding shells, $n=2,3$, can be seen as being made up of
$m_n$ identical strands, each helically wound with a linear pitch $\lambda_n$
in a cylinder of radius $r_n$.
Choosing conventionally the strand direction closest to the wire axis $z$,
we follow each strand and determine the number $M_n$ of 
periodically repeated cells and the corresponding number of turns $N_n$
required for that strand to connect back to itself. Hence
$\lambda_n = M_n c / N_n$, 
where $c=60.83\,\rm\AA$ is our cell length.
For wire B7 we find
$m_2=7$,  $r_2=3.3\,\rm\AA$, $M_2=7$, $N_2=2$, $\lambda_2 = (7/2)c$,
and
$m_3=11$, $r_3=6.0\,\rm\AA$, $M_3=11$, $N_3=7$, $\lambda_3 = (11/7)c$.
Strikingly, the two consecutive shells are not simply
commensurate, which implies in this case an overall coincidence
length of $(7\times 11)c$, or $4684\,\rm\AA$!

How should we understand these new morphologies?
The basic explanation is clearly related to surface energy.
In the weird wires, contrary to fcc wires, the ``bulk'' packing
is not so good, but the surface packing is excellent. 
Unlike crystalline wires (which must by necessity
possess different crystallographic facets), the weird wires
in fact display a {\em single}, optimal surface structure.
This feature is also a characteristic of, e.g., icosahedral
clusters, and represents the winning ingredient at very small radii.

Quantitatively, we can write the crystalline wire energy 
per atom as
\begin{equation}
E_c(R)= - E_{\rm coh} + \frac{2\Omega_c}{R} \bar\gamma
+ \frac{\Omega_c}{\pi R^2} 8\bar\mu
\label{eq:ecryst}
\end{equation}
where $E_{\rm coh}$ is the bulk cohesive energy,
$\Omega_c = \pi R^2 L/N$
($L$ is the wire length) is the average atomic volume,
$\bar\gamma$ is the surface energy (averaged on the different facets), and
$\bar\mu$ is the average edge energy per unit length
(there are 8 edges in our geometry).
$\bar\gamma$ can be written as
$\bar\gamma =  \gamma_{111} 
[\alpha_{111} + \alpha_{100} f_{100} + \alpha_{110} f_{110}]$
where $\gamma_{111}$ is the close-packed surface energy,
$\alpha_{ijk}$ is the fraction of exposed area relative to
$\langle ijk\rangle$ facets, and
$f_{ijk} = \gamma_{ijk}/\gamma_{111}$.
Moreover, $8\bar\mu = 4\mu_1 + 4\mu_2$
where $\mu_1$ and $\mu_2$ are the edge energies relative to the 
$\langle 111\rangle$--$\langle 100\rangle$ and
$\langle 111\rangle$--$\langle 110\rangle$ edge respectively.
The weird wire energy, on the other hand, is much simpler:
\begin{equation}
E_w(R)= - E_{\rm coh} + \Delta + \frac{2\Omega_w}{R} \gamma_{111}
\label{eq:eweird}
\end{equation}
reflecting a) a single smooth closed-packed surface and b) no edges.
This surface gain is countered by bad packing, causing
a bulk energy increase $\Delta$ and an average atomic volume
$\Omega_w$ (defined as for the crystalline wires) slightly
larger than $\Omega_c$. 
The crystalline-weird transition will take place at
\begin{equation}
\frac{1}{R_c} =
\frac{1}{R_{c\circ}} \left[
\sqrt{ 1 + \left( \frac{\Omega_c \delta}{R_{c\circ} \Delta} \right)^2 }
- \frac{\Omega_c \delta}{R_{c\circ} \Delta} \right]
\label{eq:transition}
\end{equation}
where $R_{c\circ}=\sqrt{8\Omega_c\bar\mu/\pi\Delta}$, and
$\delta = \bar\gamma - (\Omega_w/\Omega_c)\gamma_{111}$
is an effective surface energy difference taking into account the different
atomic volumes, that can be
either positive or negative, depending on the material.
From separate calculations, we know {\em a priori}
all parameters, except for $\Delta$ and $\bar\mu$.
The geometries of the sections of fcc wires are adequately
represented by
$\alpha_{111} = 0.55$, $\alpha_{100} = 0.25$
and $\alpha_{110} = 0.20$.
For Al (Pb),
$\Omega_c = 16.1$ (27.9) $\rm \AA^3$,
$\Omega_w/\Omega_c = 1.10$ (1.00),
$\gamma_{111} = 54.3$ (37.5) $\rm meV/\AA^2$,
$f_{100} = 1.08$ (1.01),
$f_{110} = 1.19$ (1.11),
$\bar\gamma = 57.5$ (38.5) $\rm meV/\AA^2$.

Figure \ref{fig:EvsR} shows the resulting plots for
eq.\ (\ref{eq:ecryst}), obtained with $\bar\mu$ as a fit parameter,
and of (\ref{eq:eweird}) with $\Delta$ as the sole parameter.
The accord with energies obtained by simulations is quite good, yielding
$\Delta = 37$ (24) meV, 
$\bar\mu = 35$ (25) $\rm meV/\AA$, and
$R_c = 5.3$ (9.8) $\rm\AA$.
These values of $\Delta$ are in good agreement with the $T=0$
energy difference between the glassy metal (obtained by simulated quenching
of the liquid) and the fcc solid, namely 51 (29) meV.
The weird wire interior is therefore close to a glass.
The edge energies are also reasonable, since they should be close 
to half the step value, or about $\gamma d /4$, which is
39 (33) $\rm meV/\AA$.
We conclude that eq.\ (\ref{eq:transition}) should have predictive power
for other materials, provided the glass, the surface and the edge
energies can be estimated.

This physics also explains the diversity of behaviour between the
two metals Al and Pb, both fcc in bulk.
The surface anisotropy of Pb is much smaller than that of Al.
All stable Al wires display
an outer shell which is triangular---like a curved (111) surface---to
a very good degree. 
In Pb, conversely, weird wires tend to possess 
a roughly square outer layer, which can be explained with the exceptionally 
low $\gamma_{100}$ \cite{bauer}.
For instance, the pentagonal (non-icosahedral)
structure B4 with (100)-like lateral 
faces is preferred over A6 (icosahedral),
similar but with (111)-like lateral faces.
In the helical wire B7, the presence 
of a relative tilt angle between third-shell and second-shell strands,
$\theta_3-\theta_2=\tan^{-1} (2\pi r_3/\lambda_3) - 
\tan^{-1} (2\pi r_2/\lambda_2) \simeq 16.0^\circ$,
appears to realize a good approximate 
coincidence of the third (outer) shell atoms with the hollow sites of 
the second shell, which in turn  
has a roughly triangular packing (inset).
This provides an amusing case of what
might be considered ``curved surface epitaxy'' between
two otherwise incompatible 2D lattices.

Finally, we compare wires with clusters.
The non-fcc wire structures we just found for Pb are plentiful. 
Conversely, the same potential is known to stabilize 
only fcc crystalline clusters and to destabilize icosahedra, 
in agreement with experimental indications \cite{kofman},
down to the lowest size (13 atoms) \cite{lim}. 
Hence, the tendency of wires to abandon 
the bulk fcc structure is much stronger than could have been expected,
in particular stronger than in clusters.
The reason is most likely related to 
the well-known fact that icosahedral shapes require (111) faces to be 
stretched in order to form a space filling structure \cite{marks}.  
Owing to the high tensile surface
stress of the metal, this circumstance can 
disfavor icosahedral clusters against 
fcc structures, when the surface energy
anisotropy is small, as in Pb. On the other hand, 
this negative factor is absent in wires, which are
open, and thus can fully adjust along $z$. 
A second difference is that icosahedral clusters have edges,
while most weird wires have none.
In both respects, therefore, wires are not necessarily 
just a two-dimensional version of clusters, and new phenomena
can arise.

Could such unsupported metal nanowires be fabricated, 
and their possibly weird structure eventually be detected? 
Wires of over a thousand \AA{}ngstroms length
can be pulled by an STM tip \cite{frenken}. They can be expected to 
possess, at least in some sections, the ultra-small radii which 
we have addressed here.
Weird structures could be sought in field ionization 
or transmission electron microscopy (TEM) images.
Also, the electronic structure of a wire is in principle a strong function
of its atomic structure. Future study might reveal a measurable
imprint of its weird shape, if present.
More work in these directions is currently planned.

We acknowledge support from EU through 
contracts ERBCHBGCT920180, ERBCHBGCT940636 and ERBCHRXCT930342,
and from INFM (Nanowire Project).

\end{multicols}
\end{document}